\documentclass{llncs}
\usepackage{psfig}

\begin{document}

\sloppy

\pagestyle{headings}
\mainmatter

\title{Generalized Quantum Search with Parallelism}

\titlerunning{}

\author{Robert Gingrich%
\thanks{\,Supported by the President's Fund, California Institute of Technology}
\and
Colin P. Williams%
\thanks{\,Supported by the NASA/JPL Center for Integrated Space Microsystems, NASA Advanced Concepts Office, the NASA Information and Computing Research Technologies Program and Caltech's President's Fund}
\and Nicolas Cerf%
\thanks{\,Supported by Caltech's President's Fund}}

\authorrunning{R. Gingrich, C. P. Williams and N. J. Cerf}

\institute{*Department of Physics, California Institute of Technology
\\
email: gingrich@krl.caltech.edu and 
\\ {**}Jet Propulsion Laboratory, California Institute of Technology, 
\\
4800 Oak Grove Drive, Pasadena, CA 91109-8099, 
\\
email: Colin.P.Williams@jpl.nasa.gov and
\\ {***}  Ecole Polytechnique, CP 165/56, Universite Libre de Bruxelles, 50 avenue F. D. Roosevelt, B-1050 Bruxelles, Belgium\\
email: ncerf@ulb.ac.be}

\maketitle

\begin{abstract}

We generalize Grover's unstructured quantum search algorithm to enable it to use an arbitrary starting superposition and an arbitrary unitary matrix {\em simultaneously}. We derive an exact formula for the probability of the generalized Grover's algorithm succeeding after $n$ iterations. We show that the fully generalized formula reduces to the special cases considered by previous authors. We then use the generalized formula to determine the optimal strategy for using the unstructured quantum search algorithm. On average the optimal strategy is about 12\% better than the naive use of Grover's algorithm.  The speedup obtained is not dramatic but it illustrates that a hybrid use of quantum computing and classical computing techniques can yield a performance that is better than either alone. We extend the analysis to the case of a society of $k$ quantum searches acting in parallel. We derive an analytic formula that connects the degree of parallelism with the optimal strategy for $k$-parallel quantum search. We then derive the formula for the expected speed of $k$-parallel quantum search.
\end{abstract}

\section{Introduction}

The field of quantum computing has undergone a rapid growth over the past few years. Simple quantum computations have already been performed using nuclear magnetic resonance \cite{gershenfeld97,cory98,cory98b,chuang98,chuang98b,jones98} and nonlinear optics technologies \cite{chuang95,franson98}. Recently, proposals for more specialized devices that rely on quantum computing have also been made \cite{dowling98}. Such devices are far from being general-purpose computers, nevertheless, they constitute significant milestones along the road to practical quantum computing. 

In tandem with these hardware developments, there has been a parallel development of new quantum algorithms. Several important quantum algorithms are now known \cite{deutsch92,shor94,grover96,cerf98,fijany98,brassard98}.  Of particular importance is the quantum algorithm for performing unstructured quantum search discovered by Lov Grover in 1996 \cite{grover96}.  Grover's algorithm is able to find a marked item in a virtual "database" containing $N$ items in $O(\sqrt{N})$ computational steps.  In contrast, the best classical algorithm requires $O(N/2)$ steps on average, and $O(N)$ steps in the worst case.  Thus Grover's algorithm exhibits a polynomial speedup over the best classical counterpart. 

Although the Grover algorithm exhibits only a polynomial speedup, it appears to be much more versatile than the other quantum algorithms. Indeed, Grover has shown how his algorithm can be used to speed up almost any other quantum algorithm \cite{grover98b}.  More surprisingly, even search problems that contain "structure" in the form of correlations between the items searched over, often reduce to an exhaustive search amongst a reduced set of possibilities.  Recently, it was shown how Grover's algorithm can be nested to exploit such problem structure \cite{cerf98}. This is significant because NP-hard problems, which are amongst the most challenging computational problems that arise in practice, possess exactly this kind of problem structure.

In order to appreciate the full versatility of Grover's algorithm it is important to examine all the ways in which it might be generalized.  For example, whereas the original Grover algorithm was started from an equally weighted superposition of eigenstates representing all the indices of the items in the database, a natural generalization would be to consider how it performs when started from an arbitrary initial superposition instead.  This refinement is important, because if Grover's algorithm is used within some larger quantum computation, it is likely to have to work on a arbitrary starting superposition rather than a specific starting eigenstate.  Similarly, the original Grover algorithm uses a particular unitary operator, the Walsh-Hadamard operator, as the basis for a sequence of unitary operations that systematically amplifies the amplitude in the target state at the expense of the amplitude in the non-target states. However, it is now known that this is not the best choice if there is partial information as to the likely location of the target item in the database. In such a situation a different unitary operator is desirable.  Hence, it is important to understand how Grover's algorithm performs when using an arbitrary unitary operator instead of the Walsh-Hadamard operator.

Each of these refinements have been analyzed in detail {\em independently}.  Biham et al. have considered the case of an arbitrary starting superposition \cite{biham98} and Grover considered the case of an arbitrary unitary operator \cite{grover98a}.  In this paper, we present the analysis of the fully generalized Grover algorithm in which we incorporate both of these effects simultaneously. Our goal is to determine the exact analytic formula for the probability of the fully generalized Grover algorithm succeeding after $n$ iterations when there are $r$ targets amongst $N$ candidates. Having obtained this formula, we will recover the Biham et al. and Grover results as special cases.  We will then show that the optimal strategy, on average, for using the fully generalized Grover algorithm consists of measuring the memory register after about 12\% fewer iterations than are needed to obtain the maximum probability of success.  This result confirms a more restricted case reported in \cite{boyer96}. Finally, we show how to boost the success probability and reduce the required coherence time by running a society of $k$ quantum searches independently in parallel.  In particular, we derive an explicit formula connecting the degree of parallelism, i.e., $k$, to the optimal number of iterations (for each agent in the society) that minimizes the expected search cost overall. We then derive the expected cost of optimal $k$-parallel quantum search. 

\section{Grover's Algorithm}

The problem we have to solve is the following.  Given a function $f(x_i)$ on a set $\mathcal{X}$ of input states such that 
\begin{equation}
f(x_i) = \left\{ \begin{array}{ll} 1 & \mbox{if} \; x_i \; \mbox{is a target 
element} \\ 0 & \mbox{otherwise} \end{array} \right. .
\end{equation}
How do we find a target element by using the least number of calls to the 
function $f(x_i)$? In general, there might be $r$ target elements, in which case any one will suffice as the answer. 

To solve the problem using Grover's algorithm we first form a Hilbert space 
with an orthonormal 
basis element for each input $x_i \in \mathcal{X}$.  Without loss of generality, we will write the target states as $| t_i \rangle$ and the non-target states as $| l_i \rangle$. In this paper we
refer to the basis of input eigenstates as the measurement basis.  Let $N =
|\mathcal{X}|$ be the cardinality of $\mathcal{X}$.  The function call
is to be implemented by a unitary operator that acts as follows:

\begin{equation}
|x_i \rangle | y \rangle \rightarrow |x_i \rangle |y \oplus f(x_i)
\rangle
\end{equation}
where $| y \rangle$ is either  $| 0 \rangle$  or  $| 1 \rangle$.  By
acting on 

\begin{equation}
\left( \sum_{i = 1}^{N - r} l_i |l_i \rangle +  \sum_{j = 1}^{r} k_j
|t_j \rangle \right) \frac{1}{\sqrt{2}} \left( | 0 \rangle - | 1 \rangle
\right)
\end{equation}
with this operator we construct the state

\begin{equation}
\left( \sum_{i = 1}^{N - r} l_i |l_i \rangle -  \sum_{j = 1}^{r} k_j      
|t_j \rangle \right) \frac{1}{\sqrt{2}} \left( | 0 \rangle - | 1 \rangle  
\right)
\end{equation}
where the $r$ measurement basis states $|t_i \rangle$ are the target
states and the $N - r$ measurement basis states  $|l_i \rangle$ are the
non-target states.  If we now disregard the state $ \frac{1}{\sqrt{2}}
\left( | 0 \rangle - | 1 \rangle \right) $ then all we have done is to
invert the phase of the target states.  Hence, the operator we have achieved is equivalent to the operator 

\begin{equation}
1 - 2 \sum_{i = 1}^{r} |t_{i}
\rangle \langle t_{i}|.
\end{equation}
although we emphasize that it is {\em not} necessary to know what the target states are {\em a priori}. 

Next we construct the operator $Q$ defined as
\begin{equation}
Q = \left( 2 |a \rangle \langle a | - 1 \right) \left( 1 - 2 \sum_{i = 
1}^{r} |t_{i} \rangle \langle t_{i}| \right)
\end{equation} 
where $|a \rangle$ can be thought of as the averaging 
state.  Different choices of $|a \rangle$ give rise to different unitary operators for performing amplitude amplification.  In the original Grover algorithm, the state $|a \rangle$ was chosen to be 

\begin{equation}
\label{hadamard}
| a \rangle = \frac{1}{\sqrt{|\mathcal{X}|}} \sum_{x \in \mathcal{X}} | x 
\rangle .
\end{equation}  
and was obtained by applying the Walsh-Hadamard operator, $U$, to a starting eigenstate $|s \rangle$, i.e., $|a \rangle = U |s \rangle$.  Hence, the operation $2 |a \rangle \langle a | -1  $, which Grover referred to as "inversion about the average", is equivalent to 
$ -U I_{s} U^{\dagger} $ with $U$ being the Walsh-Hadamard operator and $I_{s}$ being $1 - 2 |s \rangle \langle s|$. 

By knowing more about the structure of the problem we can choose other 
vectors $|a \rangle$ that will allow us to find a target state faster.  
Techniques for doing this are given in \cite{grover98b}.

Fortunately, in order to determine what action the operator $Q$ performs, it is sufficient to focus on a two-dimensional subspace.  The basis vectors of this subspace can be written as 

\begin{equation}
\begin{array}{ll}
|t \rangle & = \frac{1}{v} \sum_{i = 1}^{r}  \langle t_{i} | a \rangle 
|t_{i} \rangle, \; \; \; \; \; v^2 =
 \sum_{i = 1}^{r} | \langle t_{i} | a \rangle |^{2}  \\
|a' \rangle & = \frac{1}{\sqrt{1 - v^2}} \left( |a \rangle - v |t \rangle
\right)
\end{array}
\end{equation}
Note that $|t \rangle$ is the normalized projection of $|a \rangle$ onto 
the space of target states and $|a' \rangle$ is the normalized 
projection of $|a \rangle$ onto the space orthogonal to $|t \rangle$.  
This choice of basis makes the calculation easiest.  The rest of the 
Hilbert space (i.e. the space orthogonal to $ |t \rangle$ and $|a' 
\rangle $ ) can be broken up into the space of target states 
($\mathcal{S}_{T}$) and non-target states ($\mathcal{S}_{L}$).  We can 
now write $Q$ as

\begin{equation}
Q = \cos \phi \left( |t \rangle \langle t| + |a' \rangle \langle a'| 
\right) + \sin \phi \left( |t \rangle \langle a'| - |a' \rangle \langle 
t| \right) + I_T - I_L , \; \; \; \; \; \phi \equiv \arccos \left[ 
1 - 2 v^2 \right]
\end{equation}
where $I_T$ and $I_L$ are the identity operators on ($\mathcal{S}_{T}$) 
and ($\mathcal{S}_{L}$) respectively.  From this we can see that $Q$ is
just a simple 
rotation matrix on $ |a' \rangle$ and $ |t \rangle$ and acts 
trivially on the rest of the space. 

An arbitrary starting superposition $|s \rangle$ for the algorithm can be written
as
\begin{equation}
|s \rangle = \alpha |t \rangle + \beta e^{ib} |a' \rangle + |\phi_{t}
\rangle + | \phi_{l} \rangle
\end{equation}  
where the states $| \phi_{t} \rangle$ and $| \phi_{l}
\rangle$ (which must have a norm less than one if the state $|s \rangle$ is to be properly normalized overall) are the components of
$|s \rangle$ in ($\mathcal{S}_{T}$) and ($\mathcal{S}_{L}$) 
respectively.  Also, $\alpha$, $\beta$ and $b$ are positive real 
numbers.  After $n$ applications of $Q$ on an arbitrary starting superposition $|s \rangle$ we have 

\begin{equation}
\begin{array}{rl}
Q^n |s \rangle = \left( \alpha \cos (n \phi) + \beta e^{i b} \sin (n \phi)
\right) |t \rangle + \left( \beta e^{i b} \cos (n \phi) - \alpha \sin (n
\phi) \right) |a' \rangle + |\phi_{t} \rangle + (- 1)^n | \phi_{l}
\rangle.
\end{array}
\end{equation}
If we measure this state our probability of success (i.e., measuring a
target state) will be given by two terms.  The first term is the magnitude
squared of $Q^n |s \rangle$ in the space $\mathcal{S}_{T}$.  This
magnitude is
$\langle  \phi_{t} |  \phi_{t} \rangle$ and is unchanged by $Q$.  The
second term is the magnitude squared of the coefficient of $|t \rangle$
which is given by 

\begin{equation}
\label{gofn}
\begin{array}{lll}
g(n) & \equiv \left| \langle t | Q^n |s \rangle \right|^2 \\
& = \left| \alpha \cos (n \phi) + \beta e^{ib} \sin (n \phi)  \right|^2 \\
& =  \frac{\alpha^2 + \beta^2}{2}  +  \frac{\alpha^2
- \beta^2}{2}  \cos (2 n \phi) + \alpha \beta \cos (b) \sin (2 n
\phi) \\
& =  \frac{\alpha^2 + \beta^2}{2} + \frac{1}{2} \left| \alpha^2 + \beta^2
e^{2ib} \right| \cos (2 n \phi + \psi) \; \; \;
\end{array}   
\end{equation}
where $\psi \equiv \arccos \left[
\frac{\beta^2 - \alpha^2}{\left| \alpha^2 + \beta^2 e^{2ib} \right|}
\right]$. This is the term that is affected by $Q$ and is the term we wish to
maximize.  The total probability of success after $n$ iterations of $Q$ acting on $|s \rangle$ is 

\begin{equation}
\label{Pofn}
p(n,r,N) = \langle  \phi_{t} |  \phi_{t} \rangle + g(n).
\end{equation}
Assuming that $n$ is continuous (an assumption that we will justify shortly) the maxima of $g(n)$, and hence the maxima of the probability of success of Grover's algorithm, are given by the following. 

\begin{equation}
\label{nsubj}
n_j = \frac{1}{2 \phi} \left( - \psi + 2 \pi j \right) \; \; 
\; \; \; j = 0, 1, 2 \ldots
\end{equation}
The value of $g(n)$ at these 
maxima is given by

\begin{equation}
g(n_j) = \frac{\alpha^2 + \beta^2}{2} + \frac{1}{2} \left| \alpha^2 + \beta^2 e^{2ib} \right|.
\end{equation}
In practice, the optimal $n$ must be an integer and typically the $n_j$'s 
are not integers.  However, since $g(n)$ can be written as 

\begin{equation}
g(n_j \pm \delta) = g(n_j) - \phi^2 \left| \alpha^2 + \beta^2 e^{2ib} \right| \delta^2 + O(\delta^4)
\end{equation}
around $n_j$ and most interesting problems will have $v \ll 1$ and hence 
$\phi \simeq 2 v \ll 1$, simply rounding $n_j$ to the nearest integer 
will not significantly change the final probability of success.  So, we have 

\begin{equation}
\label{pfinal}
p(n_{max},r,N) = \frac{\alpha^2 + \beta^2}{2} + \frac{1}{2} \left| \alpha^2 + \beta^2 e^{2ib} \right| +  \langle \phi_{t} |
\phi_{t} \rangle - O(v^2)
\end{equation}
as the probability of measuring a target state after $n_{max} = n_{j}$
applications of $Q$. 

\begin{figure}[t]
\vskip 1 cm
\centerline{\psfig{figure=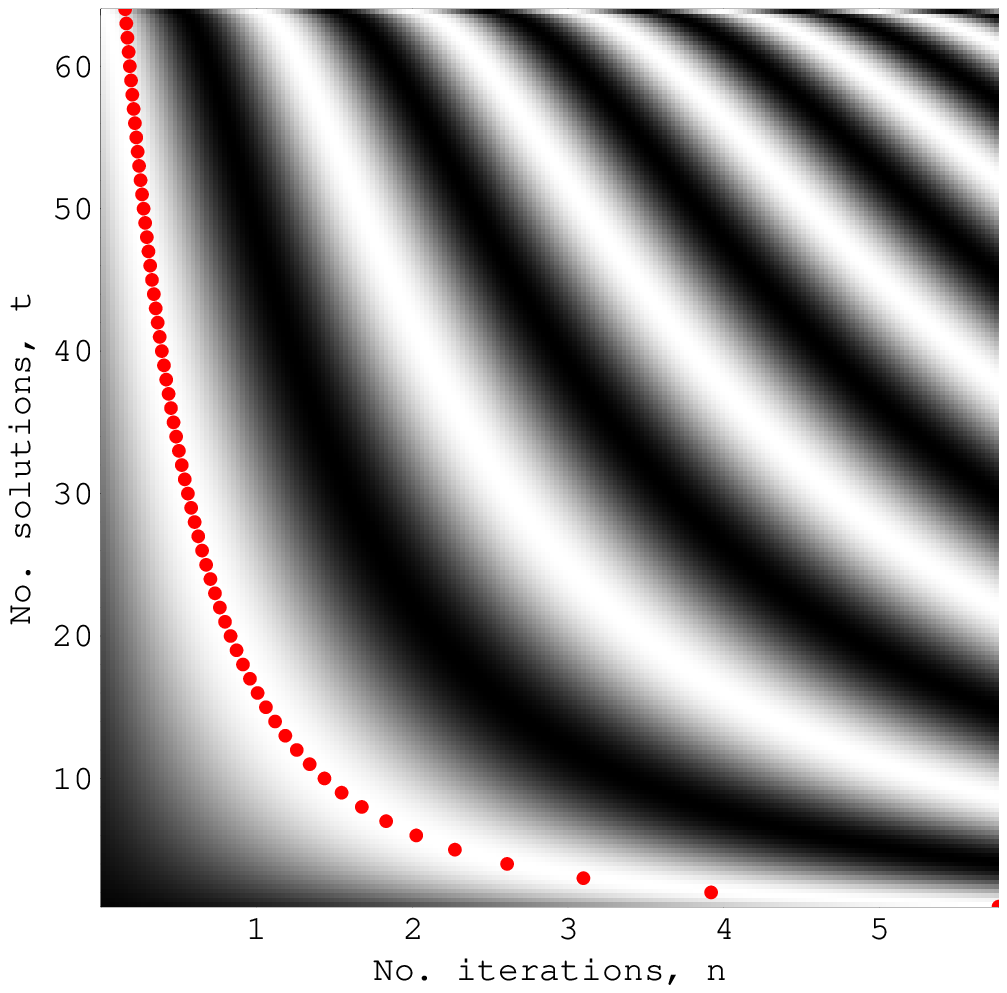,width=4.5in,angle=0}}
\label{groverFig}
\caption{Plot of the probability of success of Grover's algorithm after $n$ iterations of amplitude amplification when there are $r$ solutions amongst $N=64$ possibilities. White regions correspond to probability 1, black regions correspond to probability 0. Notice the periodicity in the success probability as the number of iterations grows.}
\vskip 0 cm
\end{figure}

\section{Recovering the Special Cases}
As a check on our fully generalized formula for the probability of success after $n$ iterations, we attempt to recover the corresponding formulae obtained in the analyses of Biham et al (for a fixed unitary operator and an arbitrary starting superposition) \cite{biham98} and Grover (for an arbitrary unitary operator and a fixed starting superposition) \cite{grover98a}.

In the case of Biham et al., the starting state is arbitrary but the averaging state $|a \rangle$ is given
by 

\begin{equation}
| a \rangle = \frac{1}{\sqrt{N}} \sum_{x \in \mathcal{X}} | x
\rangle
\end{equation}
In this case

\begin{equation}
\begin{array}{ll}
v & = \sqrt{\frac{r}{N}} \\
|t \rangle &  = \frac{1}{\sqrt{r}} \sum_{i = 1}^{r} |t_i \rangle \\
|a' \rangle &  = \frac{1}{\sqrt{N - r}} \sum_{i = 1}^{N - r} |l_i \rangle
\end{array}
\end{equation}
In the
analysis of \cite{biham98} they use $\overline{k}(0)$ and $\overline{l}(0)$ to
represent the average amplitudes, in $|s \rangle$, of the target and non-target states
respectively, and $\sigma_k$ and $\sigma_l$ to represent the standard deviations of
those amplitudes.  With some algebra one can see that the following relationships connect our notation to theirs:

\begin{equation}
\begin{array}{rl}
\alpha &  \longrightarrow \overline{k}(0)  \sqrt{r} \\
\beta e^{ib} &  \longrightarrow \overline{l}(0) \sqrt{N - r} \\
\langle \phi_{t} | \phi_{t} \rangle &  \longrightarrow r \sigma_{l}^2  \\
\langle \phi_{l} | \phi_{l} \rangle &  \longrightarrow (N - r) \sigma_{k}^2  \\
\phi & \longrightarrow \omega \\
\psi &  \longrightarrow 2 \mbox{Re} [\phi] - \pi \\
n &  \longrightarrow t\\ 
n_0 &  \longrightarrow T.
\end{array}
\end{equation}
If you substitute these relationships into equations~\ref{gofn},~\ref{nsubj},
and~\ref{pfinal} you will reproduce the
results of \cite{biham98}.  

The second special case, in which $|a \rangle$ (the averaging state) is an unknown normalized vector while $|s
\rangle$ is given by 

\begin{equation}
\begin{array}{rl}
| s \rangle & = |a \rangle \\
& = \sqrt{1 - v^2} |a' \rangle + v | t \rangle.
\end{array}
\end{equation}
was considered by Grover. Hence, $\alpha = v$, $\beta = \sqrt{1 - v^2}$ and $b = 0$.  Also,
$ | \phi_{t} \rangle = | \phi_{l} \rangle = 0$.  These substitutions lead to $\psi = \phi$.  Plugging
this into equations~\ref{nsubj} and~\ref{pfinal} we get

\begin{equation}
\begin{array}{rl}
n_{max} & = \frac{\pi}{2 \phi} -  \frac{1}{2} \\
    & = \frac{\pi}{4 v} - \frac{1}{2} - \frac{\pi v}{24} + O(v^2)
\end{array}
\end{equation}
and

\begin{equation}
p(n_{max}) = 1 - O(v^2)
\end{equation} 
which agree with the results of the paper. If we examine
equation \ref{Pofn} in this case we get

\begin{equation}
p(n)  = \frac{1 - \cos [ (2 n + 1) \phi ] }{2}
\end{equation}
as the probability of measuring a target state after $n$ iterations of
$Q$.  

\section{Application of the Formula for $p(n)$} 
Next, we show how to apply our analytic formula for the probability of success after $n$ iterations, $P(n)$, to slightly speed up the quantum unstructured search algorithm.  Although the speedup we obtain is not dramatic, it is worth making the point that it is possible at all as Christoph Zalka has proved, correctly, that Grover's algorithm is exactly optimal \cite{zalka98}.  Many people have assumed, therefore, that it is impossible to beat Grover's algorithm.  However, by combining techniques of quantum computing with those of classical computing we show that it is possible to do a little bit better than Grover's algorithm on average.  The result we report was apparently discovered previously by Boyer et al. \cite{boyer96}  It is shown here to persist for the case of fully generalized quantum search. 

We consider a punctuated quantum search algorithm that works as follows: 
\\
\\
\textbf{Algorithm: Punctuated Quantum Search}\\
1. Run the quantum search algorithm for $n$ iterations.\\
2. Read the memory register.\\
3. If the result is a target state halt, else reset the register to the starting superposition and return to step 1.\\

The average time, $T_{avg}(n)$,
it will take to find a target state if we stop the generalized quantum search algorithm after $n$ iterations of $Q$ is

\begin{equation}
\label{avgCost1}
\begin{array}{rl}
T_{avg}(n) & = \sum_{i = 1}^{\infty} \left( 1 - p(n) \right)^{i - 1} p(n) i n \\
& = \frac{n}{p(n)} \\
& = \frac{2 n}{1 - \cos[(2 n + 1) \phi]}.
\end{array}
\end{equation}
We can find the optimal strategy, i.e., the best number of iterations to use before we attempt to measure the register, by minimizing the expected running time $T_{avg}$.  To do this, we set the derivative of $T_{avg}$ to zero and solve for $n = n_{opt}$. 

\begin{equation}
\frac{2 - 2 \cos [(2 n + 1) \phi] - 4 n \phi \sin [(2n + 1) \phi]}{\left(
1 - \cos[(2 n + 1) \phi] \right)^2} = 0.
\end{equation}
Typically $n$ will be much larger than one so we can make the
approximation $(2 n + 1) \phi \simeq 2 n \phi \equiv x$.  By dropping the denominator we obtain

\begin{equation}
\begin{array}{rl}
1 - \cos x & = x \sin x \\
2 \sin^2 \left( \frac{x}{2} \right) & = 2 x \sin \left( \frac{x}{2} \right) \cos \left( \frac{x}{2} \right) \\
x & = \tan \left( \frac{x}{2} \right) . 
\end{array} 
\end{equation}
which gives $x = 2.3311$ as the lowest positive solution.  This solution corresponds to
the minimum of the function.  Hence
the optimal value of $n$ is 

\begin{equation}
n_{opt} \simeq \frac{1.1655}{\phi}
\end{equation}
This value of $n$ corresponds to an average number of iterations of 

\begin{equation}
T_{avg}(n = n_{opt}) \simeq \frac{1.3801}{\phi}
\end{equation}
compared to $\frac{1.5708}{\phi}$ iterations if we run Grover's algorithm
until
the probability is maximal.  

It is interesting to note that if we restrict the analysis some more to
the case where $|a \rangle$ is given by equation \ref{hadamard}, and there
is only one target state then $T_{avg} \left( n_{opt}
\right) \simeq 0.6900 \sqrt{N}$.  This is faster than the lower bounds in
\cite{grover98c}, \cite{bennett97}, \cite{boyer96}, and \cite{zalka98}, but we are using
a somewhat different model.  They are looking at the minimum time it would take
without measuring to find a solution with certainty up to errors
from rounding $n_{max}$ to the nearest integer.  The model we use allows for punctuated measurements and resets of the quantum search algorithm. Nevertheless, the punctuated quantum search algorithm is faster on average.  Note that we have assumed that the time it takes to measure and reset the algorithm is
negligible. This is reasonable as it only requires one function
call.

The punctuated quantum search algorithm has another advantage in that it
should make it easier to eliminate decoherence.  If we wait until we have
the maximal probability of measuring a target state then we must maintain
coherence for $\frac{1.5708}{\phi}$ steps as compared to only
$\frac{1.1655}{\phi}$ steps for the fastest measure and restart method. In
fact if we are willing to settle for an average time equal to the time it
takes to have maximal probability then coherence need only be maintained
for $\frac{0.7854}{\phi}$ steps at a time.  

\section{$k$-Parallel Quantum Search} 
A way to speed up Grover's algorithm still further is to have a society of $k$ computational agents all running Grover's algorithm independently at the same time.  This is promising because the standard deviation

\begin{equation}
\sigma_T = \frac{n}{p(n)} \sqrt{[1 - p(n)] [1 - p(n) + p(n)^2]}
\end{equation}
of the method we have already analyzed is fairly large and hence having
multiple algorithms running may give a considerable speed up. 

Suppose that we know that there are exactly $r$ solutions amongst $N$ candidates.  Given $p(n, r, N)$, the probability of success for a single agent after $n$ iterations, we can boost the success probability by using $k$ agents acting in parallel. In particular, the probability that at least one agent, in a society of $k$ independent agents, succeeds after each agent has undergone $n$ iterations is given by
\begin{equation}
p_k(n,r,N) = 1 - (1 - p(n,r,N))^k
\end{equation}
Thus the expected cost, $T_{avg}^{(k)}$, of performing $k$-parallel quantum search is given by
\begin{equation}
T_{avg}^{(k)} = \sum_{j = 1}^{\infty} {j~n~p_k(n,r,N) (1 - p_k(n,r,N))^{j-1}} \\
= {n \over p_k(n,r,N)} \\
= {n \over {1-\left(\cos(\frac{1}{2} (1+2n) \arccos(1-\frac{2r}{N})\right)^{2k}}}
\end{equation}
As in equation \ref{avgCost1} we can find the value of $n$ that minimizes the expected cost. For $ {{r} \over {N}} \ll 1$, i.e., when there are very few solutions amongst the items searched over, $\arccos(1 - {{2r} \over {N}}) \approx 2 \sqrt{{r} \over {N}}$. Hence the average cost for $k$-parallel quantum search is given by:

\begin{equation}
T_{avg}^{(k)}(n,r,N)~\approx~{{n} \over {1-\left(\cos\left((1+2n) \sqrt{{r} \over {N}}\right)\right)^{2k}}}
\end{equation}

To find the mimimum, we find where $\frac{\partial{T_{avg}^{(k)}(n,r,N)}}{\partial{n}}$ is equal to zero. This derivative is given by:

\begin{equation}
\frac{\partial{T_{avg}^{(k)}(n,r,N)}}{\partial{n}}~=~{{1 - (\cos((1+2n) \sqrt{{r} \over {N}}))^{2k} \left(1 + 4 k n \sqrt{{r} \over {N}} \tan((1+2n) \sqrt{{r} \over {N}}) \right)} \over {\left( -1+\left(\cos((1+2n)\sqrt{{r} \over {N}})\right)^{2k}\right)^2}}
\end{equation}

Substituting $x=(1+2n) \sqrt{{r} \over {N}}$ and realizing that $n \gg 1$ we obtain the following:

\begin{equation}
\label{derivInx}
\frac{\partial{T_{avg}^{(k)}(n,r,N)}}{\partial{n}}~\approx~{{1 - (\cos(x))^{2k} \left(1 + 2 k x \tan(x) \right)} \over {\left( -1+\left(\cos(x)\right)^{2k}\right)^2}}
\end{equation}

The variable $x<1$ provided $n< \frac{1}{2}\left( \sqrt{{N}\over{r}}-1\right)$. We know that we can solve the problem with near certainty if we iterate Grover's algorithm to the maximum probability state in $O(\frac{\pi}{4}\sqrt{{N}\over{r}})$ iterations.  Hence, for a large enough number of parallel search agents, $k$, there is a reasonable chance that the optimum number of iterations, $n_{optimum}(r,N,k)$ at which the expected search cost is minimized, satisfies the criterion that $x<1$. We therefore expand equation \ref{derivInx} as a series approximation in $x$ about $x=0$ to order $O(x^2)$. Hence

\begin{equation}
\frac{\partial{T_{avg}^{(k)}(n,r,N)}}{\partial{n}}~\approx~{{-60+10(-1+3k)x^2+3(-1+5k^2)x^4}\over{60 k x^2}}
\end{equation}

As this equation is fourth order in $x$ it can be solved analytically. Three of the roots are non-physical but one corresponds to an approximation to the true minimum of $T_{avg}^{(k)}(n,r,N)$. Specifically, we find that $\frac{\partial{T_{avg}^{(k)}(n,r,N)}}{\partial{n}}=0$ and $T_{avg}^{(k)}(n,r,N)$ is minimized when $x$ is given by
\begin{equation}
x~=~\sqrt{{{5-15k+\sqrt{5} \sqrt{-31-30k+225k^2}}\over{-3+15k^2}}}\end{equation}

We note that $x<1$ for all $k\geq2$. Hence, the derivation of the optimum formula is self-consistent. Hence, as $x=(1+2n) \sqrt{{r} \over {N}}$, we obtain the formula for $n_{optimal}(r,N,k)$, the predicted optimal number of iterations to use for each of $k$ quantum searches acting independently in parallel as
\begin{equation}
n_{optimal}(r,N,k)~\approx~{{1}\over{2}}{ \left( \sqrt{{{5-15k+\sqrt{5} \sqrt{-31-30k+225k^2}}\over{-3+15k^2}}} \sqrt{{N}\over{r}} - 1\right)}
\end{equation}


\begin{figure}[t]
\vskip 1 cm
\centerline{\psfig{figure=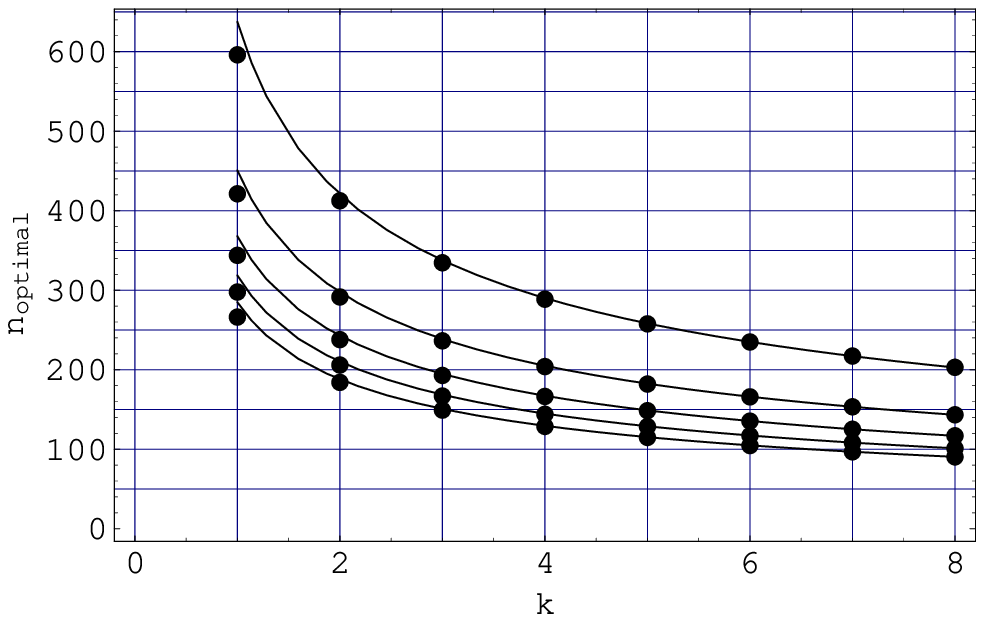,width=4.5in,angle=0}}
\label{nOptimalFig}
\caption{Plot of the optimal number of iterations to use in $k$-parallel quantum search as a function of the degree of parallelism $k$ for $r=1$ to $r=5$ solutions (top to bottom in the figure) for the case of a database of size $N=2^{20}$. The solid curves are produced by numerical optimization. The points are given by our approximate formula for $n_{optimal}(r,N,k)$}
\vskip 0 cm
\end{figure}

Hence the expected cost for optimal $k$-parallel quantum search is given explicitly by:
\begin{equation}
T_{avg}^{(k)}(n,r,N)~\approx~{{{  \sqrt{{{5-15k+\sqrt{5} \sqrt{-31-30k+225k^2}}\over{-3+15k^2}}} \sqrt{{N}\over{r}} - 1}} \over {2-2\cos^{2k}\left( \sqrt{{{5-15k+\sqrt{5} \sqrt{-31-30k+225k^2}}\over{-3+15k^2}}}  \right)}}
\end{equation}

\section{Conclusions}
In this paper we have shown how to generalize the analysis of unstructured quantum search to incorporate the effects of an arbitrary starting superposition and an arbitrary unitary operator (or, equivalently, arbitrary averaging state) {\em simultaneously}. We have also shown that, rather than iterating the amplitude amplification operator until the maximum probability of success state it attained (i.e., for $O(0.785398 \sqrt{N})$ iterations) it is better to stop after only $O(0.6900 \sqrt{N})$ iterations (i.e., 88\% of the maximum probability case). This strategy, is therefore approximately 12\% faster than Grover's algorithm on average.

Moreover, an ever better quantum search algorithm can be obtained by running $k$ independent quantum searches in parallel, stopping as soon as any of the quantum searches finds a solution. We find that the optimal $k$-parallel punctuated quantum search strategy is different from that of single agent punctuated quantum search strategy. In general, the higher the degree of classical parallelism the less (parallel) time is needed to perform the quantum computation.  This intuition is captured analytically in equation (38), which gives the explicit connection between the number of amplitude amplification iterations as a function of the degree of parallelism $k$. This result is of practical utility to experimentalists. In particular, in any physical embodiment of a quantum search there will be some natural coherence time beyond which the computation becomes unreliable. Of course, quantum error correction allows this time to be extended greatly, arguably indefinitively, if the individual error probability per gate operation can be made sufficiently small. While we believe this to be true, in practice it might be extraordinarily difficult to achieve. Instead, if we can predict the degree of parallelism needed so that the quantum search has a good chance of completing within the natural coherence time of the physical system being used as the quantum computer, then the strategy of massive parallelism might provide a realistic alternative to relying solely on quantum error correction. Thus we see the classical parallelism as an adjunct to quantum error correction rather than a replacement for it. Equation (38) exposes precisely the space/time tradeoff between quantum coherent computing and classical parallelism, at least in the context of unstructured quantum search.

\end{document}